\documentclass[preprint,pteplogo]{ptephy_v2}

\usepackage{amsmath}
\allowdisplaybreaks
\usepackage{amssymb}
\usepackage{graphicx}
\usepackage{dcolumn}
\usepackage[shortlabels]{enumitem}
\usepackage{longtable}
\usepackage[svgnames]{xcolor}
\usepackage{ulem} %%%% for ``uwave''
\usepackage{mathrsfs}
\usepackage{bm}
\usepackage{mathtools}

\usepackage{newtxtext,newtxmath}
\usepackage{hyperref}

\AtBeginDocument{\mathcode`v=\varv}
\AtBeginDocument{\mathcode`w=\varw}

\newcommand{\iu}{{\rm i}}

\begin{document}

\title{ Field-theoretical approach to
neutral pion contribution to muon $g-2$ }

\author{ Masashi Hayakawa }
\affil{
Department of Physics, Nagoya University, Nagoya, 464-8602 Japan
%\email{xxxx@xxxx.ac.jp}
}
\affil{
Nishina Center, RIKEN, Wako, 351-0198 Japan
}

%\author{Insert last author name here\thanks{These authors contributed equally to this work}}
%\affil{Insert last author address here}

%%% To include the collaborator name... Please use the command "\collaborator"
%%% For example: \collaborator{ATLAS Collaboration}

\begin{abstract}%
 The hadronic light-by-light scattering induces 
a substantial contribution
to the muon $g-2$ with present accuracy of its measurement. 
 The effect caused by the neutral pion through this scattering
on the muon $g-2$ has been scrutinized intensively by model calculation
and lattice QCD simulation.
 All of those estimates have been done based on 
one formula, but it has one pathological aspect.
 The purpose of this article is to resolve such a pathology
from a field-theoretical approach, {\it i.e.}
by defining the effect in question  
as the neutral pion intermediate state contribution to 
the QCD vacuum expectation value of time-ordered product of
four electromagnetic currents. 
 It leads 
(1) the pion pole term corresponding
to the imaginary part of the pion propagator
$\displaystyle{\frac{1}{q^2 - m_\pi^2 + \mathrm{i} \epsilon}}$, 
(2) and no principle value term
corresponding to its real part, 
resolving the pathology.
% If the result captures the essence of the neutral pion effect, 
%its actual size may be smaller than the present consensus value.
%its actual size may differ from the known consensus value.
% Another direction to approximate
%the hadronic light-by-light scattering 
%without the neutral pion intermediate state contribution
%is also pointed out.
\end{abstract}

\subjectindex{muon, magnetic dipole moment, QCD}

\maketitle

\section{Introduction}
\label{sec:intro}
 The hadronic light-by-light scattering contribution to the muon
$g-2$ ($a_\mu$) 
is one of non-negligible QCD corrections 
to confront the measured value of $a_\mu$ 
with the standard model prediction $a_\mu(\mathrm{th})$.
 Its first study dates back to Ref.~\cite{Kinoshita:1984it}.
 A single neutral pion, 
which is a one-particle state in pure QCD, {\it i.e.}
stable if the other known interactions are switched off, 
induces scattering of two photons.
 Ref.~\cite{Kinoshita:1984it}
attempted to describe this scattering process 
by the low-energy effective Lagrangian containing neutral pion field. 
 The leading order relevant interaction is given by 
Wess-Zumino-Witten Lagrangian.
 Its use as two $\pi^0 \gamma(^*) \gamma(^*)$-vertices
($\gamma^*$ means ``off-shell'' photon at the moment)
connected by the pion propagator
leads the hadronic light-by-light scattering amplitude, 
but its contribution to the muon $g-2$ is logarithmically divergent.
 Nevertheless, such a point-like interaction never reflects
the actual behavior of the $\pi^0 \gamma \gamma^*$-vertex
for the photon with high energy.
 Taking their damping behavior into account 
is quite crucial to draw a definite prediction on 
the neutral pion contribution to the muon $g-2$ through 
the light-by-light scattering, 
refereed to as the pion pole contribution. 

 Closer examination of the neutral pion contribution 
and the others such as the charged pion loop contribution 
showed \cite{Hayakawa:1995ps,Hayakawa:1996ki,Bijnens:1995cc,Bijnens:1995xf}
that the neutral pion is likely the main source
of the hadronic light-by-light scattering contribution to 
the muon $g-2$.
 Numerous number of studies have been done since then
to get more reliable prediction on this contribution.
 Most of them has examined how much the result is affected 
by the postulates on the $\pi^0 \gamma^* \gamma^*$-vertex 
proposed in progress of the experiments on $\pi^0 \gamma \gamma^*$.
 The latest lattice QCD simulation 
provides further refinement of the study 
on the neutral pion pole contribution
\cite{Gerardin:2016cqj,Gerardin:2019vio,
Gerardin:2023naa,ExtendedTwistedMass:2023hin,Lin:2024khg}.

 Although the lattice calculation has been attempted  
for the full hadronic light-by-light scattering contribution
\cite{Blum:2019ugy,Blum:2023vlm,Chao:2021tvp}, 
it is still a challenging subject.
 Thus, the traditional approach, which has been called as
the dispersive analysis by Ref.~\cite{Colangelo:2014dfa}, 
will continue to play an important role to estimate
the hadronic light-by-light scattering contribution to the muon $g-2$.

 To explain the motivation of this paper,
we recall the well-known expression for the pion pole contribution
($k_{(4)} \coloneq - \left(k_{(1)} + k_{(2)} + k_{(3)}\right)$)
with the pion mass $m_\pi$
\cite{Knecht:2001qf,Colangelo:2017urn}
\begin{align}
 &
 - \Pi^{\nu_{(1)} \nu_{(2)} \nu_{(3)} \nu}
  \left(k_{(1)},\,k_{(2)},\,k_{(3)}\right)_{\pi^0}
  \nonumber \\
 &\ 
 =
 \quad 
  \varepsilon^{\nu_{(1)} \nu_{(2)} \rho \sigma}
    k_{(1),\,\rho}\,k_{(2),\,\sigma}\,
  \varepsilon^{\nu_{(3)} \nu \kappa \tau} k_{(3),\,\kappa}\,k_{(4),\,\tau}
  \,F\left(k_{(1)}^2\,,\,k_{(2)}^2\right) F\left(k_{(3)}^2\,,\,k_{(4)}^2\right)
  \nonumber \\
 &\qquad
 \times
 \frac{1}{\left(k_{(1)} + k_{(2)}\right)^2 - m_\pi^2 + i \epsilon}
   \nonumber \\
 &\quad
 +
  \varepsilon^{\nu_{(2)} \nu_{(3)} \rho \sigma}
    k_{(2),\,\rho}\,k_{(3),\,\sigma}\,
  \varepsilon^{\nu_{(1)} \nu \kappa \tau} k_{(1),\,\kappa}\,k_{(4),\,\tau}
  \,F\left(k_{(2)}^2\,,\,k_{(3)}^2\right) F\left(k_{(1)}^2\,,\,k_{(4)}^2\right)
  \nonumber \\
 &\qquad
 \times
 \frac{1}{\left(k_{(2)} + k_{(3)}\right)^2 - m_\pi^2 + i \epsilon}
  \nonumber \\
 &\quad
 +
  \varepsilon^{\nu_{(3)} \nu_{(1)} \rho \sigma}
    k_{(3),\,\rho}\,k_{(1),\,\sigma}\,
  \varepsilon^{\nu_{(2)} \nu \kappa \tau} k_{(2),\,\kappa}\,k_{(4),\,\tau}
  \,F\left(k_{(3)}^2\,,\,k_{(1)}^2\right) F\left(k_{(2)}^2\,,\,k_{(4)}^2\right)
  \nonumber \\
 &\qquad
 \times
 \frac{1}{\left(k_{(3)} + k_{(1)}\right)^2 - m_\pi^2 + i \epsilon}
  \,. \label{eq:known-pion-pole-contribution}
\end{align}
 $F\left(k_{(1)}^2\,,\,k_{(2)}^2\right)$ in the above denotes
the transition form factor in QCD
\footnote{
 The symbol $\mathscr{T}$ reorders a sequence of the operators 
enclosed in the bracket
according to the order of their time coordinates 
as described in detail at the beginning 
of Sec.~\ref{sec:intermediateStates}.
}
\begin{align}
 &
 \left(-\iu\right)
 \varepsilon^{\nu_{(1)} \nu_{(2)} \rho \sigma} k_{(1),\,\rho}\,h_{\sigma}\,
 F\left(k_{(1)}^2\,,\,\left(h - k_{(1)}\right)^2\right)
  \nonumber \\
 &\quad
 =
 \int d^4 z_{(1)}\,e^{\iu k_{(1)} \cdot z_{(1)}}\,
 \left<0\right|
  \mathscr{T}\left[ J^{\nu_{(1)}}\left(z_{(1)}\right)\,
    J^{\nu_{(2)}}\left( 0 \right) \right]
 \left|\boldsymbol{h}\right>
  \label{eq:def:pion-formfactor}
\end{align}
for the quark electromagnetic (EM) current 
%%%
\begin{align}
 J^\mu(x) &\coloneq 
 \sum_ f Q_f\,\overline{f}(x)\,\gamma^\mu\,f(x)\,,
\end{align}
%%%
with the EM charge $Q_f$ of the quark of the flavor $f$
($\displaystyle{Q_u = \frac{2}{3}}$ for the up-quark $u$), 
and the neutral pion state $\left|\boldsymbol{h}\right>$ 
carrying the momentum $\boldsymbol{h}$ and the energy 
$h^0 
= E_{\left|\boldsymbol{h}\right|} \coloneq \sqrt{\left|\boldsymbol{h}\right|^2 + m_\pi^2}$
normalized such that
\begin{align}
 \left< \boldsymbol{h} \right| \left. \boldsymbol{h}^\prime \right>
 =
 2 E_{\left|\boldsymbol{h}\right|} \left(2\pi\right)^3 
 \delta^3 \left( \boldsymbol{h} - \boldsymbol{h}^\prime \right) \,.
 \label{eq:normalizationOfOneParticleState}
\end{align}
 All analyses of the pion pole contribution to the muon $g-2$ 
have been done based on 
a certain integral inferred from Eq.~(\ref{eq:known-pion-pole-contribution})
since the work \cite{Knecht:2001qf}.
 However, Eq.~(\ref{eq:known-pion-pole-contribution}) has a pathology.
 For instance, 
the four-momentum $\left(- k_{(1)} - k_{(2)}\right)$
carried by the pion propagator in the first term in 
Eq.~(\ref{eq:known-pion-pole-contribution}) 
is off mass shell for generic momenta $k_{(1)}$, $k_{(2)}$, 
while Eq.~(\ref{eq:def:pion-formfactor}) defines 
the transition form factor $F\left(k_{(1)}^2\,,\,k_{(2)}^2\right)$
only for on-shell pion, {\it i.e.} for $k_{(1)}\,,\,k_{(2)}$ such that
$\left( k_{(1)} + k_{(2)} \right)^2 = m_\pi^2$. 
 The expression (\ref{eq:known-pion-pole-contribution})
does not respect energy-momentum conservation for generic 
four momenta, unless $F\left(k^2\,,\,l^2\right)$
is replaced by the counterpart for the off-shell pion. 

 In practice, all of the phenomenological model calculations 
have implicitly assumed the functions extrapolated to the off-shell pion
and respected energy-momentum conservation\,:
They do not respect
Eq.~(\ref{eq:known-pion-pole-contribution}) faithfully.
 As mentioned in the beginning of this section, 
the low energy effective theory with the neutral pion 
as a dynamical variable immediately allows to find a Feynman rule 
leading some expression
with $\pi^0 \gamma^* \gamma^*$ vertex for the off-shell pion, 
which is not Eq.~(\ref{eq:known-pion-pole-contribution}) strictly.
 Moreover, it does not seem to give the ``pion pole'' contribution
for the following reason. 
 The embedding of the expression (\ref{eq:known-pion-pole-contribution})
into the formula to get the muon $g-2$
\cite{Knecht:2001qf,Colangelo:2017urn}
is done with both of 
two loop momenta circulating in three internal photon lines
connecting QCD part with the muon
Wick-rotated.
 Thus, it is quite unlikely that
the imaginary part of 
$\displaystyle{
\frac{1}{\left(k_{(1)} + k_{(2)}\right)^2 - m_\pi^2 + i \epsilon}}$
in proportion to 
$\delta\left(\left(k_{(1)} + k_{(2)}\right)^2 - m_\pi^2\right)$
contributes to the muon $g-2$ to be estimated in that way
(See Sec.~\ref{sec:conclusion} for more detail on this point.).
%the value obtained from Eq.~(\ref{eq:known-pion-pole-contribution}).
% The same calculation becomes, in fact, impossible 
%if we try to estimate the pion pole contribution 
%according to the fully data-driven approach
%from lattice QCD study on pion transition form factor, 
%because it is unable to define the off-shell pion state field-theoretically. 

 The only literature 
that presents description of how to derive the formula
(\ref{eq:known-pion-pole-contribution})
would be Ref.~\cite{Colangelo:2017urn}.
 The derivation is, however, not satisfactory :
\begin{itemize}
\item 
 Ref.~\cite{Colangelo:2017urn} first obtains the imaginary part of
one of the three terms in Eq.~(\ref{eq:known-pion-pole-contribution}).
 To do so, they deal with the matrix element
with two off-shell photons in the initial and the final states, 
$\left|\gamma^*(k_{(1)})\,\gamma^*(k_{(2)})\right>$.
 However, such states do not exist in QCD.
 It is thus impossible to use unitarity
of the S matrix $\mathbb{S} = \mathbb{I} + \iu \mathbb{T}$ in QCD, 
$\mathbb{I} = \mathbb{S}^\dagger \mathbb{S}$, 
that is, we cannot sandwich that equation by
$\left<\gamma^*(-k_{(1)})\,\gamma^*(-k_{(2)})\right|$
and $\left|\gamma^*(k_{(3)})\,\gamma^*(k_{(4)})\right>$.
 In ($\mathrm{QCD} + \mathrm{QED}$), the situation is worse:
in addition to no notion of off-shell photon, 
%in which a neutral pion no longer exists as a one-particle state 
%because it decays through $\pi^0 \rightarrow \gamma \gamma$. 
a neutral pion no longer exists
as a single particle state of the system.
%as a pole 
%of the spectral function of the Green function of
%two pseudoscalar density operators
%$P(x) \coloneqq 
%\overline{u}(x)\,\iu\,\gamma_5\,u(x) - 
%\overline{d}(x)\,\iu\,\gamma_5\,d(x)$.
%%
\item
 The form (\ref{eq:known-pion-pole-contribution}) with the pion propagator
seems to be just postulated.
\item
 Ref.~\cite{Colangelo:2017urn} explains
that the coefficient multiplied to the pion propagator
can be derived from its imaginary part.
 However, the imaginary part is just on the pion's mass shell 
so that it is impossible to get full information on
the coefficient which in general depends on the momentum carried 
by the pion off mass shell.
\item
 The off-shell pion is hard to define in QCD unambiguously.
\end{itemize}
 It is thus worthwhile scrutinizing the validity of
Eq.~(\ref{eq:known-pion-pole-contribution})
from the scratch, 
and finding an alternative formula 
representing the neutral pion contribution
to the hadronic light-by-light scattering amplitude if necessary.

 As in the lattice QCD simulation 
\cite{Blum:2019ugy,Blum:2023vlm,Chao:2021tvp}, 
the hadronic light-by-light scattering amplitude
$\Pi^{\nu_{(1)} \nu_{(2)} \nu_{(3)} \mu}
  \left( k_{(1)}\,,\,k_{(2)}\,,\,k_{(3)} \right)$
in the momentum space
is defined here by the connected Green function of four $J^\mu(x)$ in QCD
%%%
\begin{align}
 &
 \iu\,\Pi^{\nu_{(1)} \nu_{(2)} \nu_{(3)} \mu}
  \left( k_{(1)}\,,\,k_{(2)}\,,\,k_{(3)} \right)
  \nonumber \\
 &
 \ \coloneq
 \int d^4 z_{(1)}\,e^{- \iu k_{(1)} \cdot z_{(1)}}
 \int d^4 z_{(2)}\,e^{- \iu k_{(2)} \cdot z_{(2)}}
 \int d^4 z_{(3)}\,e^{- \iu k_{(3)} \cdot z_{(3)}}
  \nonumber \\
 &
 \qquad \times
 \left<0\right|
  \mathscr{T}
  \left[
   J^{\nu_{(1)}}\left(z_{(1)}\right)\,J^{\nu_{(2)}}\left(z_{(2)}\right)
   J^{\nu_{(3)}}\left(z_{(3)}\right)\,J^{\mu}\left(0\right)
 \right]
 \left|0\right>_{\mathrm{conn}}
  \,. \label{eq:def:HLbLamp}
\end{align}
%%%
 Here, all momenta $k_{(1)}$, $k_{(2)}$, $k_{(3)}$
are defined to flow into the interacting region.
 Note that Eq.~(\ref{eq:def:HLbLamp}) does not refer to
any off-shell photon states.
 The only possible way that can be done in quantum field theory is
to define the effect of the neutral pion 
on this hadronic light-by-light scattering amplitude
by singling out the contribution of neutral pion intermediate state.
 The main result (\ref{eq:our-pion-pole-contribution}) of this paper is that
the neutral pion intermediate state contribution, 
or the neutral pion contribution in short defined as such,  
is actually given just in terms of the physical pion transition form factor
$F\left(k_{(1)}^2\,,\,k_{(2)}^2\right)$.
 There does not exist the contribution corresponding to 
the principal value from
$\displaystyle{
\frac{1}{\left(k_{(1)} + k_{(2)}\right)^2 - m_\pi^2 + i \epsilon}}$
in Eq.~(\ref{eq:known-pion-pole-contribution}).

 The approach of this work is unable to deduce
the overall normalization $\beta$ for the neutral pion contribution
(\ref{eq:our-pion-pole-contribution}).
 Here, the low energy effective theory is adopted as 
a guiding principle to fix it\,:$\beta = 1$.
% The determination of in Eq.~(\ref{eq:our-pion-pole-contribution})
%requires some guiding principle that cannot be found here. 
 All of the details leading the result will be presented
for the reader to take a critical look at this work in what follows.

% No useful expression to estimate 
%the neutral pion contribution to the muon $g-2$ has not been obtained yet.
% Thus, the detail analysis on how much the known consensus value
%for the this contribution is affected by the conclusion of this work
%must be left as the future work.

\section{Insertion of intermediate states into time-ordered product}
\label{sec:intermediateStates}

 For the purpose of getting used to 
the time-ordered product of four EM currents, 
it is written down explicitly.
 First, here are eight cases
that will be relevant to the $s$-channels 
after insertion of a complete basis 
$\left\{\left|v\right>\right\}$
of QCD states in the middle of
four EM currents
\footnote{
 The subscript ``$\mathrm{conn}$'' will be omitted in what follows.
 To avoid complication of presentation, 
we are insensitive to the equal-time cases.
}
\,:
\begin{align}
 &
 \left< 0 \right|
  \mathscr{T}
  \left[
   J^{\nu_{(1)}}(z_{(1)})\,J^{\nu_{(2)}}(z_{(2)})
   \,J^{\nu_{(3)}}(z_{(3)})\,J^{\nu}(0)
  \right]
 \left| 0 \right>
  \nonumber \\
% 1
 &\quad
 =
 \left< 0 \right|
  J^{\nu_{(1)}}(z_{(1)})\,J^{\nu_{(2)}}(z_{(2)})
  \,J^{\nu_{(3)}}(z_{(3)})\,J^{\nu}(0)
 \left| 0 \right> \quad
 \left( z_{(1)}^0 > z_{(2)}^0 > z_{(3)}^0 > 0 \right)
  \nonumber \\
% 2
 &\quad
 =
 \left< 0 \right|
  J^{\nu_{(1)}}(z_{(1)})\,J^{\nu_{(2)}}(z_{(2)})
  \,J^{\nu}(0)\,J^{\nu_{(3)}}(z_{(3)})
 \left| 0 \right> \quad
 \left( z_{(1)}^0 > z_{(2)}^0 > 0 > z_{(3)}^0 \right)
  \nonumber \\
% 3
 &\quad
 =
 \left< 0 \right|
  J^{\nu_{(2)}}(z_{(2)})\,J^{\nu_{(1)}}(z_{(1)})
  \,J^{\nu_{(3)}}(z_{(3)})\,J^{\nu}(0)
 \left| 0 \right> \quad
 \left( z_{(2)}^0 > z_{(1)}^0 > z_{(3)}^0 > 0 \right)
  \nonumber \\
% 4
 &\quad
 =
 \left< 0 \right|
  J^{\nu_{(2)}}(z_{(2)})\,J^{\nu_{(1)}}(z_{(1)})
  \,J^{\nu}(0)\,J^{\nu_{(3)}}(z_{(3)})
 \left| 0 \right> \quad
 \left( z_{(2)}^0 > z_{(1)}^0 > 0 > z_{(3)}^0 \right)
  \nonumber \\
% 5
 &\quad
 =
 \left< 0 \right|
  J^{\nu_{(3)}}(z_{(3)})\,J^{\nu}(0)
  \,J^{\nu_{(1)}}(z_{(1)})\,J^{\nu_{(2)}}(z_{(2)})
 \left| 0 \right> \quad
 \left( z_{(3)}^0 > 0 > z_{(1)}^0 > z_{(2)}^0 \right)
  \nonumber \\
% 6
 &\quad
 =
 \left< 0 \right|
  J^{\nu_{(3)}}(z_{(3)})\,J^{\nu}(0)
  \,J^{\nu_{(2)}}(z_{(2)})\,J^{\nu_{(1)}}(z_{(1)})
 \left| 0 \right> \quad
 \left( z_{(3)}^0 > 0 > z_{(2)}^0 > z_{(1)}^0 \right)
  \nonumber \\
% 7
 &\quad
 =
 \left< 0 \right|
  J^{\nu}(0)\,J^{\nu_{(3)}}(z_{(3)})
  \,J^{\nu_{(1)}}(z_{(1)})\,J^{\nu_{(2)}}(z_{(2)})
 \left| 0 \right> \quad
 \left( 0 > z_{(3)}^0 > z_{(1)}^0 > z_{(2)}^0 \right)
  \nonumber \\
% 8
 &\quad
 =
 \left< 0 \right|
  J^{\nu}(0)\,J^{\nu_{(3)}}(z_{(3)})
  \,J^{\nu_{(2)}}(z_{(2)})\,J^{\nu_{(1)}}(z_{(1)})
 \left| 0 \right> \quad
 \left( 0 > z_{(3)}^0 > z_{(2)}^0 > z_{(1)}^0 \right)
  \,. \label{eq:time-order:s}
\end{align} 

 Here are eight cases relevant to $u$-channels\,:
\begin{align}
 &
 \left< 0 \right|
  \mathscr{T}
  \left[
   J^{\nu_{(1)}}(z_{(1)})\,J^{\nu_{(2)}}(z_{(2)})
   \,J^{\nu_{(3)}}(z_{(3)})\,J^{\nu}(0)
  \right]
 \left| 0 \right>
  \nonumber \\
% 9
 &\quad
 =
 \left< 0 \right|
  J^{\nu_{(2)}}(z_{(2)})\,J^{\nu_{(3)}}(z_{(3)})
  \,J^{\nu_{(1)}}(z_{(1)})\,J^{\nu}(0)
 \left| 0 \right> \quad
 \left( z_{(2)}^0 > z_{(3)}^0 > z_{(1)}^0 > 0 \right)
  \nonumber \\
% 10
 &\quad
 =
 \left< 0 \right|
  J^{\nu_{(2)}}(z_{(2)})\,J^{\nu_{(3)}}(z_{(3)})
  \,J^{\nu}(0)\,J^{\nu_{(1)}}(z_{(1)})
 \left| 0 \right> \quad
 \left( z_{(2)}^0 > z_{(3)}^0 > 0 > z_{(1)}^0 \right)
  \nonumber \\
% 11
 &\quad
 =
 \left< 0 \right|
  J^{\nu_{(3)}}(z_{(3)})\,J^{\nu_{(2)}}(z_{(2)})
  \,J^{\nu_{(1)}}(z_{(1)})\,J^{\nu}(0)
 \left| 0 \right> \quad
 \left( z_{(3)}^0 > z_{(2)}^0 > z_{(1)}^0 > 0 \right)
  \nonumber \\
% 12
 &\quad
 =
 \left< 0 \right|
  J^{\nu_{(3)}}(z_{(3)})\,J^{\nu_{(2)}}(z_{(2)})
  \,J^{\nu}(0)\,J^{\nu_{(1)}}(z_{(1)})
 \left| 0 \right> \quad
 \left( z_{(3)}^0 > z_{(2)}^0 > 0 > z_{(1)}^0 \right)
  \nonumber \\
% 13
 &\quad
 =
 \left< 0 \right|
  J^{\nu_{(1)}}(z_{(1)})\,J^{\nu}(0)
  \,J^{\nu_{(2)}}(z_{(2)})\,J^{\nu_{(3)}}(z_{(3)})
 \left| 0 \right> \quad
 \left( z_{(1)}^0 > 0 > z_{(2)}^0 > z_{(3)}^0 \right)
  \nonumber \\
% 14
 &\quad
 =
 \left< 0 \right|
  J^{\nu_{(1)}}(z_{(1)})\,J^{\nu}(0)
  \,J^{\nu_{(3)}}(z_{(3)})\,J^{\nu_{(2)}}(z_{(2)})
 \left| 0 \right> \quad
 \left( z_{(1)}^0 > 0 > z_{(3)}^0 > z_{(2)}^0 \right)
  \nonumber \\
% 15
 &\quad
 =
 \left< 0 \right|
  J^{\nu}(0)\,J^{\nu_{(1)}}(z_{(1)})
  \,J^{\nu_{(2)}}(z_{(2)})\,J^{\nu_{(3)}}(z_{(3)}) 
\left| 0 \right> \quad
 \left( 0 > z_{(1)}^0 > z_{(2)}^0 > z_{(3)}^0 \right)
  \nonumber \\
% 16
 &\quad
 =
 \left< 0 \right|
  J^{\nu}(0)\,J^{\nu_{(1)}}(z_{(1)})
  \,J^{\nu_{(3)}}(z_{(3)})\,J^{\nu_{(2)}}(z_{(2)})
 \left| 0 \right> \quad
 \left( 0 > z_{(1)}^0 > z_{(3)}^0 > z_{(2)}^0 \right)
  \,.\label{eq:time-order:u}
\end{align}
 Finally, the following are eight cases relevant to 
the $t$-channels\,:
\begin{align}
 &
 \left< 0 \right|
  \mathscr{T}
  \left[
   J^{\nu_{(1)}}(z_{(1)})\,J^{\nu_{(2)}}(z_{(2)})
   \,J^{\nu_{(3)}}(z_{(3)})\,J^{\nu}(0)
  \right]
 \left| 0 \right>
  \nonumber \\
% 17
 &\quad
 =
 \left< 0 \right|
  J^{\nu_{(3)}}(z_{(3)})\,J^{\nu_{(1)}}(z_{(1)})
  \,J^{\nu_{(2)}}(z_{(2)})\,J^{\nu}(0)
 \left| 0 \right> \quad
 \left( z_{(3)}^0 > z_{(1)}^0 > z_{(2)}^0 > 0 \right)
  \nonumber \\
% 18
 &\quad
 =
 \left< 0 \right|
  J^{\nu_{(3)}}(z_{(3)})\,J^{\nu_{(1)}}(z_{(1)})
  \,J^{\nu}(0)\,J^{\nu_{(2)}}(z_{(2)})
 \left| 0 \right> \quad
 \left( z_{(3)}^0 > z_{(1)}^0 > 0 > z_{(2)}^0 \right)
  \nonumber \\
% 19
 &\quad
 =
 \left< 0 \right|
  J^{\nu_{(1)}}(z_{(1)})\,J^{\nu_{(3)}}(z_{(3)})
  \,J^{\nu_{(2)}}(z_{(2)})\,J^{\nu}(0)
 \left| 0 \right> \quad
 \left( z_{(1)}^0 > z_{(3)}^0 > z_{(2)}^0 > 0 \right)
  \nonumber \\
% 20
 &\quad
 =
 \left< 0 \right|
  J^{\nu_{(1)}}(z_{(1)})\,J^{\nu_{(3)}}(z_{(3)})
  \,J^{\nu}(0)\,J^{\nu_{(2)}}(z_{(2)})
 \left| 0 \right> \quad
 \left( z_{(1)}^0 > z_{(3)}^0 > 0 > z_{(2)}^0 \right)
  \nonumber \\
% 21
 &\quad
 =
 \left< 0 \right|
  J^{\nu_{(2)}}(z_{(2)})\,J^{\nu}(0)
  \,J^{\nu_{(3)}}(z_{(3)})\,J^{\nu_{(1)}}(z_{(1)})
 \left| 0 \right> \quad
 \left( z_{(2)}^0 > 0 > z_{(3)}^0 > z_{(1)}^0 \right)
  \nonumber \\
% 22
 &\quad
 =
 \left< 0 \right|
  J^{\nu_{(2)}}(z_{(2)})\,J^{\nu}(0)
  \,J^{\nu_{(1)}}(z_{(1)})\,J^{\nu_{(3)}}(z_{(3)})
 \left| 0 \right> \quad
 \left( z_{(2)}^0 > 0 > z_{(1)}^0 > z_{(3)}^0 \right)
  \nonumber \\
% 23
 &\quad
 =
 \left< 0 \right|
  J^{\nu}(0)\,J^{\nu_{(2)}}(z_{(2)})
  \,J^{\nu_{(3)}}(z_{(3)})\,J^{\nu_{(1)}}(z_{(1)})
 \left| 0 \right> \quad
 \left( 0 > z_{(2)}^0 > z_{(3)}^0 > z_{(1)}^0 \right)
  \nonumber \\
% 24
 &\quad
 =
 \left< 0 \right|
  J^{\nu}(0)\,J^{\nu_{(2)}}(z_{(2)})
  \,J^{\nu_{(1)}}(z_{(1)})\,J^{\nu_{(3)}}(z_{(3)})
 \left| 0 \right> \quad
 \left( 0 > z_{(2)}^0 > z_{(1)}^0 > z_{(3)}^0 \right)
  \,.\label{eq:time-order:t}
\end{align}

 Because direct analytical study of 
$\left< 0 \right|
  \mathscr{T}
  \left[
   J^{\nu_{(1)}}(z_{(1)})\,J^{\nu_{(2)}}(z_{(2)})
   \,J^{\nu_{(3)}}(z_{(3)})\,J^{\nu}(0)
  \right] 
 \left| 0 \right>_{\mathrm{conn}}$ is intractable at present, 
we are inclined to insert a complete basis of QCD states
and to seek a chance to approximate it by
estimating the dominant intermediate state contributions.
 In this section, we go back to examination of possible ways to do so\,:
\begin{enumerate}
\item 
\ For the vacuum expectation value of four EM current operators
corresponding to each order of four time coordinates, 
there are three possible ways to insert a complete
set $\left\{\left| v \right>\right\}$ of 
orthonormalized QCD states
(See Eq.~(\ref{eq:normalizationOfOneParticleState})
for the normalization of the one-particle state). 
 For instance, for $z_{(1)}^0 > z_{(2)}^0 > z_{(3)}^0 > 0$, 
\begin{align}
 &
 \left< 0 \right|
  J^{\nu_{(1)}}(z_{(1)})\,J^{\nu_{(2)}}(z_{(2)})
  \,J^{\nu_{(3)}}(z_{(3)})\,J^{\nu}(0)
 \left| 0 \right>
  \nonumber \\
 &\quad
 =
 \sum_v
 \left< 0 \right|
  J^{\nu_{(1)}}(z_{(1)})\,J^{\nu_{(2)}}(z_{(2)})
 \left| v \right>
 \left< v \right|
  \,J^{\nu_{(3)}}(z_{(3)})\,J^{\nu}(0)
 \left| 0 \right>
  \,, \nonumber \\
 &\quad
 =
 \sum_v
 \left< 0 \right|
  J^{\nu_{(1)}}(z_{(1)})\,J^{\nu_{(2)}}(z_{(2)})
  \,J^{\nu_{(3)}}(z_{(3)})
 \left| v \right>
 \left< v \right| J^{\nu}(0) \left| 0 \right>
  \,, \nonumber \\
 &\quad
 =
 \sum_v
 \left< 0 \right|
  J^{\nu_{(1)}}(z_{(1)})
 \left| v \right>
 \left< v \right|
  J^{\nu_{(2)}}(z_{(2)})\,J^{\nu_{(3)}}(z_{(3)})\,J^{\nu}(0)
 \left| 0 \right>
  \,. \label{eq:ex:insertionOfStates}
\end{align}
 For instance, the second way of insertion in the above contains
the matrix element
$\left< v \right| J^{\nu}(0) \left| 0 \right>$ that determines
the purely QCD vacuum polarization contribution to the muon $g-2$, 
{\it i.e.} the one relevant to 
the cross section to produce the hadronic state $\left|v\right>$
by $e^+ e^-$ collision at the leading order of QED.
\item
\ If the matrix elements corresponding to all of the intermediate states
can be obtained and summed over exactly, 
each equality in Eq.~(\ref{eq:ex:insertionOfStates}) holds exactly.
 Because it is impossible to do so in practice,
we choose one among three of Eq.~(\ref{eq:ex:insertionOfStates})
and try to find an optimal method to approximate it
by estimating the matrix elements for a limited number of 
intermediate states $\left|v\right>$.
 To draw one estimate of the hadronic light-by-light scattering 
contribution, 
we cannot sum  
$ \left< 0 \right|
  J^{\nu_{(1)}}(z_{(1)})\,J^{\nu_{(2)}}(z_{(2)})
  \,J^{\nu_{(3)}}(z_{(3)})
 \left| v \right>
 \left< v \right| J^{\nu}(0) \left| 0 \right>$
and 
$ \left< 0 \right|
  J^{\nu_{(1)}}(z_{(1)})\,J^{\nu_{(2)}}(z_{(2)})
 \left| u \right>
 \left< u \right|
  \,J^{\nu_{(3)}}(z_{(3)})\,J^{\nu}(0)
 \left| 0 \right>
$.
%%%
\item
\ {\it In principle}, 
we can make {\it different} choices for the way of insertion
of a complete set of states for $24$ orders of four time coordinates.
 This may be meaningful if we have some method to calculate 
the relevant matrix elements of {\it non-time-ordered} product 
of the current operators. 
 In practice, this is not the case.
 We are thus inclined to 
choose {\it only one way} of insertion
of a complete set of the hadronic states common
to {\it all of $24$ orders} of four time coordinates, 
to find out the expression 
written in terms of %the transition amplitudes, namely, 
the matrix elements of the {\it time-ordered} product of
the current operators, which are expected to be accessible by
the experiments or the lattice QCD simulation.
\end{enumerate}

 With the above general remarks in our mind, 
let's pay our attention to
the first way in Eq.~(\ref{eq:ex:insertionOfStates}) 
as the insertion of the intermediate states
for all of $24$ cases in Eqs.~(\ref{eq:time-order:s}), 
(\ref{eq:time-order:u}) and (\ref{eq:time-order:t}).
 The detail analysis yields the expression 
given in terms of the matrix elements of 
the time-ordered product of two current operators
\begin{align}
 &
 \left< 0 \right|
  \mathscr{T}
  \left[
   J^{\nu_{(1)}}(z_{(1)})\,J^{\nu_{(2)}}(z_{(2)})\,J^{\nu_{(3)}}(z_{(3)})\,
   J^{\nu}(0)
  \right] 
 \left| 0 \right>
 %\left| 0 \right>_{\rm QCD}
  \nonumber \\
 &\quad
 =
 \ 
 \theta\left(
  \min\left(z_{(1)}^0\,,\,z_{(2)}^0\right)
  -
  \max\left(z_{(3)}^0\,,\,0\right)
 \right)
  \nonumber \\
 &\qquad
 \times
 \sum_v
  \left< 0 \right|
   \mathscr{T} \left[ J^{\nu_{(1)}}(z_{(1)})\,J^{\nu_{(2)}}(z_{(2)}) \right]
  \left| v \right>
  \left< v \right|
   \mathscr{T} \left[ J^{\nu_{(3)}}(z_{(3)})\,J^{\nu}(0) \right] 
  \left| 0 \right>
  \nonumber \\
% 2
 &\quad
 +
 \theta\left(
  \min\left(z_{(3)}^0\,,\,0\right)
  -
  \max\left(z_{(1)}^0\,,\,z_{(2)}^0\right)
 \right)
  \nonumber \\
 &\qquad
 \times
 \sum_v
  \left< 0 \right|
   \mathscr{T}\left[ J^{\nu_{(3)}}(z_{(3)})\,J^{\nu}(0) \right]
  \left| v \right>
  \left< v \right|
   \mathscr{T}\left[ J^{\nu_{(1)}}(z_{(1)})\,J^{\nu_{(2)}}(z_{(2)}) \right]
  \left| 0 \right>
   \nonumber \\
% 3
 &
 \quad
 +
 \theta\left(
  \min\left(z_{(2)}^0\,,\,z_{(3)}^0\right)
  -
  \max\left(z_{(1)}^0\,,\,0\right)
 \right)
  \nonumber \\
 &\qquad
 \times
 \sum_v
  \left< 0 \right|
   \mathscr{T}\left[ J^{\nu_{(2)}}(z_{(2)})\,J^{\nu_{(3)}}(z_{(3)}) \right]
  \left| v \right>
  \left< v \right|
   \mathscr{T}\left[ J^{\nu_{(1)}}(z_{(1)})\,J^{\nu}(0) \right]
  \left| 0 \right>
  \nonumber \\
% 4
 &\quad
 +
 \theta\left(
  \min\left(z_{(1)}^0\,,\,0\right)
  -
  \max\left(z_{(2)}^0\,,\,z_{(3)}^0\right)
 \right)
  \nonumber \\
 &\qquad
 \times
 \sum_v
  \left< 0 \right|
   \mathscr{T}\left[ J^{\nu_{(1)}}(z_{(1)})\,J^{\nu}(0) \right]
  \left| v \right>
  \left< v \right|
   \mathscr{T}\left[ J^{\nu_{(2)}}(z_{(2)})\,J^{\nu_{(3)}}(z_{(3)}) \right]
  \left| 0 \right>
   \nonumber \\
% 5
 &
 \quad
 +
 \theta\left(
  \min\left(z_{(3)}^0\,,\,z_{(1)}^0\right)
  -
  \max\left(z_{(2)}^0\,,\,0\right)
 \right)
  \nonumber \\
 &\qquad
 \times
 \sum_v
  \left< 0 \right|
   \mathscr{T}\left[ J^{\nu_{(3)}}(z_{(3)})\,J^{\nu_{(1)}}(z_{(1)}) \right]
  \left| v \right>
  \left< v \right|
   \mathscr{T}\left[ J^{\nu_{(2)}}(z_{(2)})\,J^{\nu}(0) \right]
  \left| 0 \right>
  \nonumber \\
% 6
 &\quad
 +
 \theta\left(
  \min\left(z_{(2)}^0\,,\,0\right)
  -
  \max\left(z_{(3)}^0\,,\,z_{(1)}^0\right)
 \right)
  \nonumber \\
 &\qquad
 \times
 \sum_v
  \left< 0 \right|
   \mathscr{T}\left[ J^{\nu_{(2)}}(z_{(2)})\,J^{\nu}(0) \right]
  \left| v \right>
  \left< v \right|
   \mathscr{T}\left[ J^{\nu_{(3)}}(z_{(3)})\,J^{\nu_{(1)}}(z_{(1)}) \right] 
  \left| 0 \right>
   \,.
  \label{eq:insert:JJ-JJ}
\end{align}
 Remember that the sum runs over the states other 
than the QCD vacuum $\left|0\right>$
because the quantity on the left hand side is the connected Green function.
 For instance, 
the first term of the right-hand side in Eq.~(\ref{eq:insert:JJ-JJ})
is obtained 
by collecting the first four cases in Eq.~(\ref{eq:time-order:s}) 
as follows.
 First, a product of the step functions
corresponding to the order $z_{(1)}^0 > z_{(2)}^0 > z_{(3)}^0 > 0$
can be written as
\begin{align}
 & 
 \theta\left(z_{(1)}^0 - z_{(2)}^0\right)\,
 \theta\left(z_{(2)}^0 - z_{(3)}^0\right)\,
 \theta\left(z_{(3)}^0\right)
  \nonumber \\
 &\ 
 =
 \left(
  \theta\left(z_{(1)}^0 - z_{(2)}^0\right)\,
  \theta\left(z_{(3)}^0\right)\,
  \theta\left(z_{(2)}^0 - z_{(3)}^0\right)
 \right)
 \theta\left(z_{(1)}^0 - z_{(2)}^0\right)\,
 \theta\left(z_{(3)}^0\right)
  \nonumber \\
 &\ 
 =
 \theta\left(
  \min\left(z_{(1)}^0\,,\,z_{(2)}^0\right)
  -
  \max\left(z_{(3)}^0\,,\,0\right)
 \right)
 \theta\left(z_{(1)}^0 - z_{(2)}^0\right)\,
 \theta\left(z_{(3)}^0\right)
  \,. \label{eq:calc:step-functions}
\end{align}
 Four products of step functions 
corresponding to the orders, 
(a) $z_{(1)}^0 > z_{(2)}^0 > z_{(3)}^0 > 0$, 
(b) $z_{(1)}^0 > z_{(2)}^0 > 0 > z_{(3)}^0$,
(c) $z_{(2)}^0 > z_{(1)}^0 > z_{(3)}^0 > 0$ and
(d) $z_{(2)}^0 > z_{(1)}^0 > 0  > z_{(3)}^0$, 
turn out to share a common step function, 
$\theta\left(
  \min\left(z_{(1)}^0\,,\,z_{(2)}^0\right)
  -
  \max\left(z_{(3)}^0\,,\,0\right)
 \right)$,
which demands
that both of $z_{(1)}^0$ and $z_{(2)}^0$ should be larger than
$z_{(3)}^0$ and $0$.
 For succeeding evaluation of the integrals, 
it should be kept in our mind that 
the region compatible with 
the condition $\min\left(z_{(1)}^0\,,\,z_{(2)}^0\right)
  -
  \max\left(z_{(3)}^0\,,\,0\right) > 0$
is the same as the union of regions specified by (a), (b), (c) and (d), 
because the union of the regions (a) and (c) is almost equivalent to 
the one with
$z_{(1)}^0 > z_{(3)}^0$, $z_{(2)}^0 > z_{(3)}^0$ and $z_{(3)} > 0$, 
and the union of the regions (b) and (d) is almost equivalent to 
the one with 
$z_{(1)}^0 > 0$, $z_{(2)}^0 > 0$ and $0 > z_{(3)}$.
 Now, the first four cases in Eq.~(\ref{eq:time-order:s}) 
yield the first term in Eq.~(\ref{eq:insert:JJ-JJ})
\begin{align}
 & 
 \theta\left(
  \min\left(z_{(1)}^0\,,\,z_{(2)}^0\right)
  -
  \max\left(z_{(3)}^0\,,\,0\right)
 \right)
  \nonumber \\
 &\ 
 \times
 \left\{
  \ \ 
  \theta\left(z_{(1)}^0 - z_{(2)}^0\right)\,
  \theta\left(z_{(3)}^0\right)
 \right.
  \nonumber \\
 &\qquad\ 
 \times
  \sum_v
   \left< 0 \right|
    J^{\nu_{(1)}}(z_{(1)})\,J^{\nu_{(2)}}(z_{(2)})
   \left| v \right>
   \left< v \right|
    J^{\nu_{(3)}}(z_{(3)})\,J^{\nu}(0)
   \left| 0 \right>
   \nonumber \\
 &\quad 
  +
  \theta\left(z_{(1)}^0 - z_{(2)}^0\right)\,
  \theta\left(-z_{(3)}^0\right)
  \nonumber \\
 &\qquad\ 
 \times
 \sum_v
  \left< 0 \right|
   J^{\nu_{(1)}}(z_{(1)})\,J^{\nu_{(2)}}(z_{(2)})
  \left| v \right>
  \left< v \right|
   J^{\nu}(0)\,J^{\nu_{(3)}}(z_{(3)})
  \left| 0 \right>
   \nonumber \\
 &\quad
  +
  \theta\left(z_{(2)}^0 - z_{(1)}^0\right)\,
  \theta\left(z_{(3)}^0\right)
  \nonumber \\
 &\qquad\ 
  \times
  \sum_v
   \left< 0 \right|
    J^{\nu_{(2)}}(z_{(2)})\,J^{\nu_{(1)}}(z_{(1)})
   \left| v \right>
   \left< v \right|
    J^{\nu_{(3)}}(z_{(3)})\,J^{\nu}(0)
   \left| 0 \right>
  \nonumber \\
 &\quad 
  +
  \theta\left(z_{(2)}^0 - z_{(1)}^0\right)\,
  \theta\left(-z_{(3)}^0\right)
  \nonumber \\
 &\qquad
 \left.
  \times
  \sum_v
   \left< 0 \right|
    J^{\nu_{(2)}}(z_{(2)})\,J^{\nu_{(1)}}(z_{(1)})
   \left| v \right>
   \left< v \right|
    J^{\nu}(0)\,J^{\nu_{(3)}}(z_{(3)})
   \left| 0 \right>
 \right\}
    \nonumber \\
%%%%%
 &\ 
 =
 \theta\left(
  \min\left(z_{(1)}^0\,,\,z_{(2)}^0\right)
  -
  \max\left(z_{(3)}^0\,,\,0\right)
 \right)
  \nonumber \\
 &\quad
 \times
 \sum_v
 \left\{
  \ \ 
  \theta\left(z_{(1)}^0 - z_{(2)}^0\right)
  \left< 0 \right|
   J^{\nu_{(1)}}(z_{(1)})\,J^{\nu_{(2)}}(z_{(2)})
  \left| v \right>
 \right.
  \nonumber \\
 &\qquad\qquad
 \left.
  +\,
  \theta\left(z_{(2)}^0 - z_{(1)}^0\right)
  \left< 0 \right|
   J^{\nu_{(2)}}(z_{(2)})\,J^{\nu_{(1)}}(z_{(1)})
  \left| v \right>
 \right\}
  \nonumber \\
 &\qquad\quad
 \times
 \left\{
  \ \ \theta\left(z_{(3)}^0\right)
  \left< v \right|
   J^{\nu_{(3)}}(z_{(3)})\,J^{\nu}(0)
  \left| 0 \right>
 \right.
  \nonumber \\
 &\qquad\qquad
 \left.
  +\,
  \theta\left(-z_{(3)}^0\right)
  \left< v \right|
   J^{\nu}(0)\,J^{\nu_{(3)}}(z_{(3)})
  \left| 0 \right>
 \right\}
  \nonumber \\
%%%
 &\ 
 =
 \theta\left(
  \min\left(z_{(1)}^0\,,\,z_{(2)}^0\right)
  -
  \max\left(z_{(3)}^0\,,\,0\right)
 \right)
  \nonumber \\
 &\quad
 \times
 \sum_v
 \left< 0 \right|
  \mathscr{T}\left[ J^{\nu_{(1)}}(z_{(1)})\,J^{\nu_{(2)}}(z_{(2)}) \right]
 \left| v \right>
 \left< v \right|
  \mathscr{T}\left[ J^{\nu_{(3)}}(z_{(3)})\,J^{\nu}(0) \right]
 \left| 0 \right> \,.
\end{align}
%%

% Eq.~(\ref{eq:insert:JJ-JJ}) is the key expression leading 
%the result of this paper.
 The neutral pion intermediate state contribution 
%(\ref{eq:insert-pion:0}) 
is defined by singling out the terms of 
$v = \pi^0(\boldsymbol{h})$
with the momentum $\boldsymbol{h}$ integrated over
according to the normalization
(\ref{eq:normalizationOfOneParticleState}) of the one-particle state
from Eq.~(\ref{eq:insert:JJ-JJ})
\begin{align}
 &
 \left.
 \left< 0 \right|
  \mathscr{T}
  \left[
   J^{\nu_{(1)}}(z_{(1)})\,J^{\nu_{(2)}}(z_{(2)})\,J^{\nu_{(3)}}(z_{(3)})\,
   J^{\nu}(0)
  \right]
 \left| 0 \right>
 \right|_{\rm \pi^0}
  \nonumber \\
% 1
 &\quad
 =
 \ 
 \theta\left(
  \min\left(z_{(1)}^0\,,\,z_{(2)}^0\right)
  -
  \max\left(z_{(3)}^0\,,\,0\right)
 \right)
  \nonumber \\
 &\qquad
 \times
 \int \frac{d^3 \boldsymbol{h}}{\left(2\pi\right)^3 2 E_{\left|\boldsymbol{h}\right|}}
  \left< 0 \right|
   \mathscr{T}\left[ J^{\nu_{(1)}}(z_{(1)}) J^{\nu_{(2)}}(z_{(2)}) \right]
  \left| \boldsymbol{h} \right>
  \left< \boldsymbol{h} \right|
   \mathscr{T}\left[ J^{\nu_{(3)}}(z_{(3)}) J^{\nu}(0) \right] 
  \left| 0 \right>
  \nonumber \\
% 2
 &\quad
 +
 \theta\left(
  \min\left(z_{(3)}^0\,,\,0\right)
  -
  \max\left(z_{(1)}^0\,,\,z_{(2)}^0\right)
 \right)
  \nonumber \\
 &\qquad
 \times
 \int \frac{d^3 \boldsymbol{h}}{\left(2\pi\right)^3 2 E_{\left|\boldsymbol{h}\right|}}
  \left< 0 \right|
   \mathscr{T}\left[ J^{\nu_{(3)}}(z_{(3)}) J^{\nu}(0) \right]
  \left| \boldsymbol{h} \right>
  \left< \boldsymbol{h} \right|
   \mathscr{T}\left[ J^{\nu_{(1)}}(z_{(1)}) J^{\nu_{(2)}}(z_{(2)}) \right]
  \left| 0 \right>
   \nonumber \\
% 3
 &
 \quad
 +
 \theta\left(
  \min\left(z_{(2)}^0\,,\,z_{(3)}^0\right)
  -
  \max\left(z_{(1)}^0\,,\,0\right)
 \right)
  \nonumber \\
 &\qquad
 \times
 \int \frac{d^3 \boldsymbol{h}}{\left(2\pi\right)^3 2 E_{\left|\boldsymbol{h}\right|}}
  \left< 0 \right|
   \mathscr{T}\left[ J^{\nu_{(2)}}(z_{(2)}) J^{\nu_{(3)}}(z_{(3)}) \right]
  \left| \boldsymbol{h} \right>
  \left< \boldsymbol{h} \right|
   \mathscr{T}\left[ J^{\nu_{(1)}}(z_{(1)}) J^{\nu}(0) \right]
  \left| 0 \right>
  \nonumber \\
% 4
 &\quad
 +
 \theta\left(
  \min\left(z_{(1)}^0\,,\,0\right)
  -
  \max\left(z_{(2)}^0\,,\,z_{(3)}^0\right)
 \right)
  \nonumber \\
 &\qquad
 \times
 \int \frac{d^3 \boldsymbol{h}}{\left(2\pi\right)^3 2 E_{\left|\boldsymbol{h}\right|}}
  \left< 0 \right|
   \mathscr{T}\left[ J^{\nu_{(1)}}(z_{(1)}) J^{\nu}(0) \right]
  \left| \boldsymbol{h} \right>
  \left< \boldsymbol{h} \right|
   \mathscr{T}\left[ J^{\nu_{(2)}}(z_{(2)}) J^{\nu_{(3)}}(z_{(3)}) \right]
  \left| 0 \right>
   \nonumber \\
% 5
 &
 \quad
 +
 \theta\left(
  \min\left(z_{(3)}^0\,,\,z_{(1)}^0\right)
  -
  \max\left(z_{(2)}^0\,,\,0\right)
 \right)
  \nonumber \\
 &\qquad
 \times
 \int \frac{d^3 \boldsymbol{h}}{\left(2\pi\right)^3 2 E_{\left|\boldsymbol{h}\right|}}
  \left< 0 \right|
   \mathscr{T}\left[ J^{\nu_{(3)}}(z_{(3)}) J^{\nu_{(1)}}(z_{(1)}) \right]
  \left| \boldsymbol{h} \right>
  \left< \boldsymbol{h} \right|
   \mathscr{T}\left[ J^{\nu_{(2)}}(z_{(2)}) J^{\nu}(0) \right]
  \left| 0 \right>
  \nonumber \\
% 6
 &\quad
 +
 \theta\left(
  \min\left(z_{(2)}^0\,,\,0\right)
  -
  \max\left(z_{(3)}^0\,,\,z_{(1)}^0\right)
 \right)
  \nonumber \\
 &\qquad
 \times
 \int \frac{d^3 \boldsymbol{h}}{\left(2\pi\right)^3 2 E_{\left|\boldsymbol{h}\right|}}
  \left< 0 \right|
   \mathscr{T}\left[ J^{\nu_{(2)}}(z_{(2)}) J^{\nu}(0) \right]
  \left| \boldsymbol{h} \right>
  \left< \boldsymbol{h} \right|
   \mathscr{T}\left[ J^{\nu_{(3)}}(z_{(3)}) J^{\nu_{(1)}}(z_{(1)}) \right]
  \left| 0 \right>
   \,.
  \label{eq:insert-pion:0}
\end{align}
%%

%%%%%%%%%%%%%%%%%%%%%%%%%%%%%%%%%%%%%%%%%%%%%%%%%%%%%%%%%%%%%%%%%%
\section{ Neutral pion intermediate state contribution }
%%%%%%%%%%%%%%%%%%%%%%%%%%%%%%%%%%%%%%%%%%%%%%%%%%%%%%%%%%%%%%%%%%

 The Fourier transformation of Eq.~(\ref{eq:insert-pion:0}) 
similar to Eq.~(\ref{eq:def:HLbLamp})
will lead the neutral pion intermediate state contribution 
$\Pi^{\nu_{(1)} \nu_{(2)} \nu_{(3)} \mu}
  \left( k_{(1)}\,,\,k_{(2)}\,,\,k_{(3)} \right)_{\pi^0}$
to the hadronic light-by-light scattering amplitude
$\Pi^{\nu_{(1)} \nu_{(2)} \nu_{(3)} \mu}
  \left( k_{(1)}\,,\,k_{(2)}\,,\,k_{(3)} \right)$.
 The purpose of this section is to try to 
express such a contribution using the form factor 
$F\left(l_{(1)}^2\,,\,l_{(2)}^2\right)$
defined in Eq.~(\ref{eq:def:pion-formfactor}).
 We will see that there are three types of terms
depending on the momenta flowing into the form factors\,:
\begin{description}
\item[
$\left( 
 \left\{k_{(1)}\,,\,k_{(2)}\right\}\,,\,\left\{k_{(3)}\,,\,k_{(4)}\right\}
\right)$
] 
The Fourier transformation of the first three terms in 
Eq.~(\ref{eq:insert-pion:0}) 
will turn out to contain the form factors, 
$F\left(k_{(1)}^2\,,\,k_{(2)}^2\right)$ 
and $F\left(k_{(3)}^2\,,\,k_{(4)}^2\right)$.
%with $k_{(4)} \coloneqq - (k_{(1)} + k_{(2)} + k_{(3)})$.
%%
\item[
$\left( 
 \left\{k_{(2)}\,,\,k_{(3)}\right\}\,,\,\left\{k_{(1)}\,,\,k_{(4)}\right\}
\right)$
] 
The Fourier transform of the third and fourth terms in 
Eq.~(\ref{eq:insert-pion:0}) 
will turn out to contain the form factors, 
$F\left(k_{(2)}^2\,,\,k_{(3)}^2\right)$ 
and $F\left(k_{(1)}^2\,,\,k_{(4)}^2\right)$.
\item[
$\left( 
 \left\{k_{(3)}\,,\,k_{(1)}\right\}\,,\,\left\{k_{(2)}\,,\,k_{(4)}\right\}
\right)$
] 
The Fourier transform of the last two terms in 
Eq.~(\ref{eq:insert-pion:0}) 
will turn out to contain the transition form factors, 
$F\left(k_{(3)}^2\,,\,k_{(1)}^2\right)$ 
and $F\left(k_{(2)}^2\,,\,k_{(4)}^2\right)$.
\end{description}

 It is easy to show that 
$F\left(l_{(1)}^2\,,\,l_{(2)}^2\right)
 = F\left(l_{(2)}^2\,,\,l_{(1)}^2\right)$.
 We also have
\begin{align}
 &
 \left<0\right|
  \mathscr{T}\left[ J^{\nu_{(1)}}\left( y_{(1)} \right)\,
    J^{\nu_{(2)}}\left( y_{(2)} \right) \right]
 \left|\boldsymbol{h}\right>
  \nonumber \\
% &\quad
% =
% \left<0\right|
%  T J^{\nu_{(1)}}\left( y_{(1)} \right)\,
%    e^{\iu \widehat{P} \cdot y_{(2)}}\,
%    J^{\nu_{(2)}}\left( 0 \right)\,e^{-\iu \widehat{P} \cdot y_{(2)}}
% \left|\boldsymbol{h}\right>_{{\rm QCD}}
%  \nonumber \\
%
% &\quad
% =
% e^{-\iu h \cdot y_{(2)}}
% \left<0\right|
%  T J^{\nu_{(1)}}\left( y_{(1)} - y_{(2)} \right)\,
%    J^{\nu_{(2)}}\left( 0 \right) 
% \left|\boldsymbol{h}\right>_{{\rm QCD}}
%  \nonumber \\
%
 &\quad
 =
 \int \frac{d^4 l_{(1)}}{\left(2\pi\right)^4}\,
 e^{-\iu l_{(1)} \cdot y_{(1)}}
 \,e^{-\iu \left(h - l_{(1)}\right) \cdot y_{(2)}}
 \left(-\iu\right)
 \varepsilon^{\nu_{(1)} \nu_{(2)} \rho \sigma} l_{(1),\,\rho}\,h_{\sigma}\,
 F\left(l_{(1)}^2\,,\,\left(h - l_{(1)}\right)^2\right)
% \ \left(l_{(2)} = h - l_{(1)} \right)
  \,. \label{eq:formFactor:coordinateSpace}
\end{align}
 It is necessary to examine whether the matrix element
$\left< \boldsymbol{h} \right|
   \mathscr{T} \left[ J^{\nu_{(1)}}(z_{(1)})\,J^{\nu_{(2)}}(z_{(2)}) \right]
  \left| 0 \right>$
can be expressed by 
$F\left(k_{(1)}^2\,,\,k_{(2)}^2\right)$ in Eq.~(\ref{eq:def:pion-formfactor}).
 We try to use the time and space reversal symmetries of QCD
with vanishing $\theta$-parameter.
 Because the intrinsic parity of the neutral pion is odd, 
$\mathcal{P} \left| \boldsymbol{h} \right> 
= - \left| - \boldsymbol{h} \right>$
under the parity transformation $\mathcal{P}$.
 Under the time reversal $\mathcal{T}$, 
the pseudoscalar density operator
$P(x) \coloneqq 
\overline{u}(x)\,\iu\,\gamma_5\,u(x) - 
\overline{d}(x)\,\iu\,\gamma_5\,d(x)$
that enables to exite a single neutral pion from  the vacuum
flips its sign.
 Thus,
$\mathcal{T} \left| \boldsymbol{h} \right> = 
- \left| - \boldsymbol{h} \right>$.
% Ref.~\cite{Seng:2016pfd} adopts this sign, but they 
%say ``the neutral pion field changes sign under $T$ under our conventions
%for the pion nucleon interactions'' in page 14.
% If this sign is incorrect, it is necessary to flip the overall sign
%of the neutral pion intermediate state contribution derived below.
 From 
\begin{align}
 &
 \mathcal{T} \mathcal{P} \left< \boldsymbol{h} \right> 
 = \left| \boldsymbol{h} \right> \,,
 \quad
 \mathcal{P} \mathcal{T} \left| 0 \right> = \left| 0 \right>
 \,, \nonumber \\
 &
 J^{\nu_{(1)}}(z_{(1)})
 = - \mathcal{P} \mathcal{T} J^{\nu_{(1)}}(-z_{(1)})\,
   \mathcal{T}^{-1} \mathcal{P} ^{-1} \,,
\end{align}
we have
\begin{align}
 \left< \boldsymbol{h} \right|
   J^{\nu_{(1)}}(z_{(1)})\,J^{\nu_{(2)}}(z_{(2)}) 
 \left| 0 \right>
 &=
 \left(
  J^{\nu_{(1)}}(z_{(1)}) \left|\boldsymbol{h}\right>\,,\,
  J^{\nu_{(2)}}(z_{(2)}) \left| 0 \right>
 \right)
  \nonumber \\
 &=
 \left(
  J^{\nu_{(1)}}(z_{(1)}) \left|\boldsymbol{h}\right>\,,\,
  \mathcal{P}^\dagger \mathcal{T}^\dagger \mathcal{T} \mathcal{P}
   J^{\nu_{(2)}}(z_{(2)}) \left| 0 \right>
 \right)
  \nonumber \\
 &=
 \left(
  \mathcal{T} \mathcal{P} J^{\nu_{(1)}}(z_{(1)}) \left|\boldsymbol{h}\right>\,,\,
  \mathcal{T} \mathcal{P}
   J^{\nu_{(2)}}(z_{(2)}) \left| 0 \right>
 \right)^{*}
  \nonumber \\
 &=
 \left(
   J^{\nu_{(1)}}(-z_{(1)}) \left|\boldsymbol{h}\right>\,,\,
   J^{\nu_{(2)}}(-z_{(2)}) \left| 0 \right>
 \right)^{*}
  \nonumber \\
 &=
 \left< 0 \right|
  J^{\nu_{(2)}}(-z_{(2)})\,J^{\nu_{(1)}}(-z_{(1)}) 
 \left| \boldsymbol{h} \right>
  \,.
\end{align}
 The complex conjugate in the third equality 
is due to the fact that the time reversal $\mathcal{T}$
is anti-unitary
\footnote{
 Recall that $U^\dagger$ for the anti-unitary operator $U$ is defined by
\begin{align}
 \left( \left|u\right> \,,\, U^\dagger \left|v\right> \right)
 &=
 \left( 
  U \left|u\right> \,,\, \left|v\right> \right)^{*}
  \,.
\end{align}
}.
 Therefore, 
\begin{align}
 \left< \boldsymbol{h} \right|
   \mathscr{T}\left[ J^{\nu_{(1)}}(z_{(1)})\,J^{\nu_{(2)}}(z_{(2)}) \right]
 \left| 0 \right>
 &=
 \theta\left(z_{(1)}^0 - z_{(2)}^0\right)
 \left< \boldsymbol{h} \right|
   J^{\nu_{(1)}}(z_{(1)})\,J^{\nu_{(2)}}(z_{(2)}) 
 \left| 0 \right>
  \nonumber \\
 &\quad
 +
 \theta\left(z_{(2)}^0 - z_{(1)}^0\right)
 \left< \boldsymbol{h} \right|
   J^{\nu_{(2)}}(z_{(2)})\,J^{\nu_{(1)}}(z_{(1)}) 
 \left| 0 \right>
  \nonumber \\
 &=
% \textcolor[named]{Red}{+}
% \left\{
  \theta\left( - z_{(2)}^0 + z_{(1)}^0 \right)
  \left< 0 \right|
   J^{\nu_{(2)}}(-z_{(2)})\,J^{\nu_{(1)}}(-z_{(1)}) 
  \left| \boldsymbol{h} \right>
% \right.
  \nonumber \\
 &\quad\ 
% \left.
  +
  \theta\left( - z_{(1)}^0 + z_{(2)}^0 \right)
  \left< 0 \right|
   J^{\nu_{(1)}}(-z_{(1)})\,J^{\nu_{(2)}}(-z_{(2)}) 
  \left| \boldsymbol{h} \right>
% \right\}
  \nonumber \\
 &=
 %\textcolor[named]{Red}{+}
 \left< 0 \right|
  \mathscr{T} \left[ J^{\nu_{(1)}}(-z_{(1)})\,J^{\nu_{(2)}}(-z_{(2)}) \right]
 \left| \boldsymbol{h} \right>
  \,. 
\end{align}
 Use of Eq.~(\ref{eq:formFactor:coordinateSpace}) on the right-hand side
leads
\begin{align}
 &
 \left< \boldsymbol{h} \right|
   \mathscr{T}\left[ J^{\nu_{(1)}}(z_{(1)})\,J^{\nu_{(2)}}(z_{(2)}) \right]
 \left| 0 \right>
  \nonumber \\
% &\ 
% =
% \int \frac{d^4 l_{(1)}}{\left(2\pi\right)^4}\,
% e^{+\iu l_{(1)} \cdot z_{(1)}}
% \,e^{+\iu \left(h - l_{(1)}\right) \cdot z_{(2)}}
% %\left(\textcolor[named]{Red}{-}\iu\right)
% \left(-\iu\right)
% \varepsilon^{\nu_{(1)} \nu_{(2)} \rho \sigma} l_{(1),\,\rho}\,h_{\sigma}\,
% F\left(l_{(1)}^2\,,\,\left(h - l_{(1)}\right)^2\right)
%  \nonumber \\
%
 &\ 
 =
 \int \frac{d^4 l_{(1)}}{\left(2\pi\right)^4}\,
 e^{-\iu l_{(1)} \cdot z_{(1)}}
 \,e^{-\iu \left(- h - l_{(1)}\right) \cdot z_{(2)}}
 %\left(\textcolor[named]{Red}{+}\iu\right)
 \iu\,
 \varepsilon^{\nu_{(1)} \nu_{(2)} \rho \sigma} l_{(1),\,\rho}\,h_{\sigma}\,
 F\left(l_{(1)}^2\,,\,\left( - h - l_{(1)}\right)^2\right)
 \,. \label{eq:anotherMatrixElement:formFactor}
\end{align}
%%
%where the direction of $l_{(1)}$ is flipped in the second line.

 Using Eqs.~(\ref{eq:formFactor:coordinateSpace})
and (\ref{eq:anotherMatrixElement:formFactor}), 
we are now ready to evaluate Fourier transformation, say, 
of the first two terms in Eq.~(\ref{eq:insert-pion:0}), 
which shall correspond to the momenta flowing into the form factors
characterized by the symbol 
$\left( 
 \left\{k_{(1)}\,,\,k_{(2)}\right\}\,,\,\left\{k_{(3)}\,,\,k_{(4)}\right\}
\right)$
\begin{align}
 &
 %(\textcolor[named]{Red}{+}\iu)\,
 \iu\,
 \Pi^{\nu_{(1)} \nu_{(2)} \nu_{(3)} \nu}
  \left(k_{(1)},\,k_{(2)},\,k_{(3)}\right)_{\pi^0,\,(3)}
  \nonumber \\
 &\quad
 =
 \int \frac{d^3 \boldsymbol{h}}
           {\left(2\pi\right)^3\,2 E_{\left|\boldsymbol{h}\right|}}
 \int d^3 \boldsymbol{l}_{(1)}
 \left(
  \frac{1}{\left(2\pi\right)^3}
  \int d^3 \boldsymbol{z}_{(1)}
   e^{\iu \left(\boldsymbol{k}_{(1)} + \boldsymbol{l}_{(1)}\right) \cdot \boldsymbol{z}_{(1)}}
  = \delta^3\left(\boldsymbol{l}_{(1)} + \boldsymbol{k}_{(1)}\right)
 \right)
  \nonumber \\
 &\qquad
 \times
  \int d^3 \boldsymbol{l}_{(3)}
  \left(
  \frac{1}{\left(2\pi\right)^3}
   \int d^3 \boldsymbol{z}_{(3)}\,
    e^{\iu \left(\boldsymbol{k}_{(3)} + \boldsymbol{l}_{(3)}\right) \cdot \boldsymbol{z}_{(3)}}
    = \delta^3\left(\boldsymbol{l}_{(3)} + \boldsymbol{k}_{(3)}\right)
 \right)
  \nonumber \\
 &\qquad
 \times
 \int_{-\infty}^\infty \frac{d l_{(1)}^0}{2\pi}\,
 \int_{-\infty}^\infty \frac{d l_{(3)}^0}{2\pi}\,
 \varepsilon^{\nu_{(1)} \nu_{(2)} \rho \sigma}
   l_{(1),\,\rho}\,h_{\sigma}\,
 \varepsilon^{\nu_{(3)} \nu \kappa \tau} l_{(3),\,\kappa}\,h_{\tau}
  \nonumber \\
 &\qquad
 \times
 \int_{-\infty}^\infty d z_{(1)}^0
  e^{-\iu \left(l_{(1)}^0 + k_{(1)}^0\right) z_{(1)}^0}
 \int_{-\infty}^\infty d z_{(2)}^0
 \int_{-\infty}^\infty d z_{(3)}^0
  e^{-\iu \left(l_{(3)}^0 + k_{(3)}^0\right) z_{(3)}^0}
  \nonumber \\
 &\qquad
 \times
 \left\{
  \ \ 
  \theta\left(
   \min\left(z_{(1)}^0\,,\,z_{(2)}^0\right)
   -
   \max\left(z_{(3)}^0\,,\,0\right)
  \right)
  e^{-\iu \left( E_{\left|\boldsymbol{k}_{(1)} + \boldsymbol{k}_{(2)}\right|} - l_{(1)}^0 + k_{(2)}^0\right) z_{(2)}^0}
 \right.
  \nonumber \\
 &\qquad\qquad
  \times
  F\left(l_{(1)}^2\,,\,\left(h-l_{(1)}\right)^2\right)
  F\left(l_{(3)}^2\,,\,\left(h+l_{(3)}\right)^2\right)
   \nonumber \\
 &\qquad\qquad
  \times
  \left(
   \int d^3 \boldsymbol{z}_{(2)}\,
    e^{\iu \left(\boldsymbol{k}_{(2)} + \boldsymbol{h} - \boldsymbol{l}_{(1)}\right)
            \cdot \boldsymbol{z}_{(2)}}
   = \left(2\pi\right)^3\,
     \delta^3\left(\boldsymbol{h} - \boldsymbol{l}_{(1)} + \boldsymbol{k}_{(2)}\right)
  \right.
   \nonumber \\
 &\qquad\qquad\qquad
  \left.
   = \left(2\pi\right)^3\,
     \delta^3\left(\boldsymbol{h} + \boldsymbol{k}_{(1)} + \boldsymbol{k}_{(2)}\right)
  \right)
  \nonumber \\
 &\qquad\quad
  +\, 
  \theta\left(
   \min\left(z_{(3)}^0\,,\,0\right)
   -
   \max\left(z_{(1)}^0\,,\,z_{(2)}^0\right)
  \right)
   e^{-\iu \left( - E_{\left|\boldsymbol{k}_{(1)} + \boldsymbol{k}_{(2)}\right|} - l_{(1)}^0 + k_{(2)}^0\right) z_{(2)}^0}
  \nonumber \\
 &\qquad\qquad
  \times
  F\left(l_{(1)}^2\,,\,\left(h + l_{(1)}\right)^2\right)
  F\left(l_{(3)}^2\,,\,\left(h - l_{(3)}\right)^2\right)
   \nonumber \\
 &\qquad\qquad
  \times
  \left(
   \int d^3 \boldsymbol{z}_{(2)}\,
    e^{\iu \left(\boldsymbol{k}_{(2)} - \boldsymbol{h} - \boldsymbol{l}_{(1)}\right)
            \cdot \boldsymbol{z}_{(2)}}
   = \left(2\pi\right)^3\,
     \delta^3\left(- \boldsymbol{l}_{(1)} + \boldsymbol{k}_{(2)} - \boldsymbol{h}\right)
  \right.
   \nonumber \\
 &\qquad\qquad\qquad
 \left.
  \left.
   = \left(2\pi\right)^3\,
     \delta^3\left(\boldsymbol{k}_{(1)} + \boldsymbol{k}_{(2)} - \boldsymbol{h}\right)
  \right)
 \right\}
 \,. \label{eq:HLbL:pionPole:3}
\end{align}
 This contains the integral over the restricted domain
($\omega_{(1)} \coloneqq l_{(1)}^0 + k_{(1)}^0$\,,\, 
$\omega_{(2)} \coloneqq \pm E_{\left|\boldsymbol{k}_{(1)}
 + \boldsymbol{k}_{(2)}\right|} - l_{(1)}^0 + k_{(2)}^0$\,,\, 
$\omega_{(3)} \coloneqq l_{(3)}^0 + k_{(3)}^0$
)
\begin{align}
 I(\boldsymbol{\omega}) &\coloneqq
 \prod_{j\,=\,1}^3 \int_{-\infty}^\infty e^{-\iu \omega_{(j)} t_{(j)}}\,
 \theta\left(
  \min(t_{(1)}\,,\,t_{(2)}) - \min\left(t_{(3)}\,,\,0\right)
 \right) \,,
\end{align}
which is ambiguous as follows.
 The remark made below Eq.~(\ref{eq:calc:step-functions}) helps
to write it as 
\begin{align}
 I(\boldsymbol{\omega}) &=
 \int_0^\infty d t_{(3)}\,e^{-\iu \omega_{(3)} t_{(3)}}
 \int_{t_{(3)}}^\infty dt_{(1)}\,e^{-\iu \omega_{(1)} t_{(1)}} 
 \int_{t_{(3)}}^\infty dt_{(2)}\,e^{-\iu \omega_{(2)} t_{(2)}}
  \nonumber \\
 & 
 + 
 \int_{-\infty}^0 d t_{(3)}\,e^{-\iu \omega_{(3)} t_{(3)}}
 \int_0^\infty dt_{(1)}\,e^{-\iu \omega_{(1)} t_{(1)}} 
 \int_0^\infty dt_{(2)}\,e^{-\iu \omega_{(2)} t_{(2)}}
  \,.
\end{align}
 The ambiguity arises since no definition is found for the integral 
\begin{align}
 \int_0^\infty dt\,e^{\iu \omega t} \,.
\end{align}
 The imaginary part may be set to $0$ by dividing 
a half line into the intervals of length $2\pi \omega^{-1}$
on each of which the integral of $\sin(\omega t)$ vanishes.
 The same consideration is applied for the real part except for 
$\omega = 0$.
 As is also inferred from energy conservation, we set
\begin{align}
 I(\boldsymbol{\omega}) &=
 a\,\prod_{j\,=\,1}^3 \delta(\omega_j) 
\end{align}
with some constant $a$.
 The integration of both sides of the equation on $\boldsymbol{R}^3$ 
immediately leads
\begin{align}
 a &= \theta(0) (2\pi)^3 \,.
\end{align}
 $I(\boldsymbol{\omega})$ thus depends
on the definition of $\beta \coloneqq 2\,\theta(0)$.
 We proceed the calculation by leaving the overall
normalization $\beta$ unfixed.

 For the first term in Eq.~(\ref{eq:HLbL:pionPole:3}), 
the relations among energies imposed by the delta functions
imply that
$h = 
\left(
 E_{\left|\boldsymbol{k}_{(1)} + \boldsymbol{k}_{(2)}\right|}, 
 - \left( \boldsymbol{k}_1 + \boldsymbol{k}_2 \right)
\right)
= - \left( k_{(1)} + k_{(2)}\right) 
= k_{(3)} + k_{(4)}$ so that
\begin{align}
 \varepsilon^{\nu_{(1)} \nu_{(2)} \rho \sigma}
  l_{(1),\,\rho}\,h_{\sigma}\,
 \varepsilon^{\nu_{(3)} \nu \kappa \tau} l_{(3),\,\kappa}\,h_{\tau}
 &=
 -\,
 \varepsilon^{\nu_{(1)} \nu_{(2)} \rho \sigma}
   k_{(1),\,\rho}\,k_{(2),\,\sigma}\,
 \varepsilon^{\nu_{(3)} \nu \kappa \tau}
   k_{(3),\,\kappa}\,k_{(4),\,\tau}
  \,.
\end{align}
%%
%with $k_{(4)} \coloneq - \left(k_{(1)} + k_{(2)} + k_{(3)}\right)$
%\footnote{
% Some literatures, such as Ref.~\cite{Colangelo:2017urn}, 
%define $k_{(4)}$ with the opposite sign.
%}.
 For the second term in Eq.~(\ref{eq:HLbL:pionPole:3}), 
$h = 
\left(
 E_{\left|\boldsymbol{k}_{(1)} + \boldsymbol{k}_{(2)}\right|}, 
 \boldsymbol{k}_1 + \boldsymbol{k}_2
\right)
= k_{(1)} + k_{(2)}
= - \left( k_{(3)} + k_{(4)}\right) $ so that
\begin{align}
 \varepsilon^{\nu_{(1)} \nu_{(2)} \rho \sigma}
  l_{(1),\,\rho}\,h_{\sigma}\,
 \varepsilon^{\nu_{(3)} \nu \kappa \tau} l_{(3),\,\kappa}\,h_{\tau}
 &=
 -\,
 \varepsilon^{\nu_{(1)} \nu_{(2)} \rho \sigma}
   k_{(1),\,\rho}\,k_{(2),\,\sigma}\,
 \varepsilon^{\nu_{(3)} \nu \kappa \tau}
   k_{(3),\,\kappa}\,k_{(4),\,\tau}
  \,.
\end{align}
 Thus, Eq.~(\ref{eq:HLbL:pionPole:3}) becomes
%\footnote{
% Note that $(\textcolor[named]{Red}{-}1)$ is multipled to the both sides
%}
%%
\begin{align}
 &
 %(\textcolor[named]{Red}{-}\iu)\,
 (-\iu)\,
 \frac{1}{\beta}\,
 \Pi^{\nu_{(1)} \nu_{(2)} \nu_{(3)} \nu}
 \left(k_{(1)},\,k_{(2)},\,k_{(3)}\right)_{\pi^0,\,(3)}
  \nonumber \\
 &\quad
 =
 \pi\,
 \varepsilon^{\nu_{(1)} \nu_{(2)} \rho \sigma}
    k_{(1),\,\rho}\,k_{(2),\,\sigma}\,
 \varepsilon^{\nu_{(3)} \nu \kappa \tau} k_{(3),\,\kappa}\,k_{(4),\,\tau}
 \,F\left(k_{(1)}^2\,,\,k_{(2)}^2\right) F\left(k_{(3)}^2\,,\,k_{(4)}^2\right)
  \nonumber \\
 &\qquad
 \times 
 \frac{1}{2\,E_{\left|{\bf k_{(1)}} + \boldsymbol{k}_{(2)}\right|}}
 \left\{
  \delta\left(
   k_{(1)}^0 + k_{(2)}^0 
   + E_{\left|\boldsymbol{k}_{(1)} + \boldsymbol{k}_{(2)}\right|}
  \right)
  +
  \delta\left(
   k_{(1)}^0 + k_{(2)}^0 
   - E_{\left|\boldsymbol{k}_{(1)} + \boldsymbol{k}_{(2)}\right|}
  \right)
 \right\}
  \,. \label{eq:result:pi0:3}
\end{align}
 Remember that 
\begin{align}
 &
 \frac{1}{2\,E_{\left|{\bf k_{(1)}} + \boldsymbol{k}_{(2)}\right|}}
 \left\{
  \delta\left(
   k_{(1)}^0 + k_{(2)}^0 
   + E_{\left|\boldsymbol{k}_{(1)} + \boldsymbol{k}_{(2)}\right|}
  \right)
  +
  \delta\left(
   k_{(1)}^0 + k_{(2)}^0 
   - E_{\left|\boldsymbol{k}_{(1)} + \boldsymbol{k}_{(2)}\right|}
  \right)
 \right\}
  \nonumber \\
 &\ 
 =
 \delta\left( \left( k_{(1)}^0 + k_{(2)}^0 \right)^2 
  - \left(E_{\left|\boldsymbol{k}_{(1)} + \boldsymbol{k}_{(2)}\right|}\right)^2 \right)
  \nonumber \\
 &\ =
 \delta\left(\left(k_{(1)} + k_{(2)}\right)^2 - m_\pi^2\right)
 \,.
\end{align}

 The other two types of the terms in Eq.~(\ref{eq:insert-pion:0})
can be evaluated in the same manner.
 Hence, 
the full neutral pion intermediate state contribution is found as
\begin{align}
 &
 (-\iu)
 \frac{1}{\beta}
 \Pi^{\nu_{(1)} \nu_{(2)} \nu_{(3)} \nu}
  \left(k_{(1)},\,k_{(2)},\,k_{(3)}\right)_{\pi^0}
  \nonumber \\
 &\ 
 =
 \quad 
 \pi\,
  \varepsilon^{\nu_{(1)} \nu_{(2)} \rho \sigma}
    k_{(1),\,\rho}\,k_{(2),\,\sigma}\,
  \varepsilon^{\nu_{(3)} \nu \kappa \tau} k_{(3),\,\kappa}\,k_{(4),\,\tau}
  \,F\left(k_{(1)}^2\,,\,k_{(2)}^2\right) F\left(k_{(3)}^2\,,\,k_{(4)}^2\right)
  \nonumber \\
 &\qquad
 \times
% \frac{1}{2\,E_{\left|{\bf k_{(1)}} + \boldsymbol{k}_{(2)}\right|}} 
% \left\{
%  \delta\left(
%   k_{(1)}^0 + k_{(2)}^0 
%   + E_{\left|\boldsymbol{k}_{(1)} + \boldsymbol{k}_{(2)}\right|}
%  \right)
%  +
%  \delta\left(
%   k_{(1)}^0 + k_{(2)}^0 
%   - E_{\left|\boldsymbol{k}_{(1)} + \boldsymbol{k}_{(2)}\right|}
%  \right)
% \right\}
 \delta\left(
  \left(k_{(1)} + k_{(2)}\right)^2 - m_\pi^2
 \right)
   \nonumber \\
 &\quad
 +
 \pi\,
  \varepsilon^{\nu_{(2)} \nu_{(3)} \rho \sigma}
    k_{(2),\,\rho}\,k_{(3),\,\sigma}\,
  \varepsilon^{\nu_{(1)} \nu \kappa \tau} k_{(1),\,\kappa}\,k_{(4),\,\tau}
  \,F\left(k_{(2)}^2\,,\,k_{(3)}^2\right) F\left(k_{(1)}^2\,,\,k_{(4)}^2\right)
  \nonumber \\
 &\qquad
 \times
% \frac{1}{2\,E_{\left|{\bf k_{(2)}} + \boldsymbol{k}_{(3)}\right|}} 
% \left\{
%  \delta\left(
%   k_{(2)}^0 + k_{(3)}^0 
%   + E_{\left|\boldsymbol{k}_{(2)} + \boldsymbol{k}_{(3)}\right|}
%  \right)
%  +
%  \delta\left(
%   k_{(2)}^0 + k_{(3)}^0 
%   - E_{\left|\boldsymbol{k}_{(2)} + \boldsymbol{k}_{(3)}\right|}
%  \right)
% \right\}
 \delta\left(
  \left(k_{(2)} + k_{(3)}\right)^2 - m_\pi^2
 \right)
  \nonumber \\
 &\quad
 +
 \pi\,
  \varepsilon^{\nu_{(3)} \nu_{(1)} \rho \sigma}
    k_{(3),\,\rho}\,k_{(1),\,\sigma}\,
  \varepsilon^{\nu_{(2)} \nu \kappa \tau} k_{(2),\,\kappa}\,k_{(4),\,\tau}
  \,F\left(k_{(3)}^2\,,\,k_{(1)}^2\right) F\left(k_{(2)}^2\,,\,k_{(4)}^2\right)
  \nonumber \\
 &\qquad
 \times
% \frac{1}{2\,E_{\left|{\bf k_{(3)}} + \boldsymbol{k}_{(1)}\right|}} 
% \left\{
%  \delta\left(
%   k_{(3)}^0 + k_{(1)}^0 
%   + E_{\left|\boldsymbol{k}_{(3)} + \boldsymbol{k}_{(1)}\right|}
%  \right)
%  +
%  \delta\left(
%   k_{(3)}^0 + k_{(1)}^0 
%   - E_{\left|\boldsymbol{k}_{(3)} + \boldsymbol{k}_{(1)}\right|}
%  \right)
% \right\}
 \delta\left(
  \left(k_{(3)} + k_{(1)}\right)^2 - m_\pi^2
 \right)
  \,. \label{eq:our-pion-pole-contribution}
\end{align}
 The three contributions appearing above
correspond to those in the $s$, $u$ and $t$ channels, respectively.

%%%%%%%%%%%%%%%%%%%%%%%%%%%%%%%%%%%%%%%%%%%%%%%%%%%%%%%%%%%%
\section{Conclusion and discussion}
\label{sec:conclusion}
%%%%%%%%%%%%%%%%%%%%%%%%%%%%%%%%%%%%%%%%%%%%%%%%%%%%%%%%%%%%

 The result (\ref{eq:our-pion-pole-contribution}) of this work
reproduces the quantity
in Eq.~(\ref{eq:known-pion-pole-contribution}) 
obtained by gathering all terms each of which is
proportional to delta function in the propagator
\begin{align}
 \frac{1}{s - m_\pi^2 + \iu \epsilon} &=
 \mathrm{P}\,\frac{1}{s - m_\pi^2}
  - \iu\,\,\pi\,\delta(s - m_\pi^2) \,,
\end{align}
up to the overall factor $\beta$, 
and deserves being called the ``neutral pion pole'' contribution.
 However, Eq.~(\ref{eq:our-pion-pole-contribution}) 
does not have any terms corresponding to the principal values
in Eq.~(\ref{eq:known-pion-pole-contribution}).

 Here, we adopt the leading-order low energy effective theory of QCD
to fix $\beta$.
 As summarized in Sec.~\ref{sec:intro}, 
it gives Eq.~(\ref{eq:known-pion-pole-contribution})
with the trivial transition form factor
inferred from Wess-Zumino-Witten Lagrangian.
 The demand that Eq.~(\ref{eq:our-pion-pole-contribution}) equals
the pole term in Eq.~(\ref{eq:known-pion-pole-contribution})
in this approximation leads $\beta = 1$ 
$\left(\displaystyle{\theta(0) = \frac{1}{2}}\right)$.

 We make a remark on the neutral pion contribution to 
the muon $g-2$\,:
\begin{enumerate}
\item
\ As indicated by the form of the delta functions
in Eq.~(\ref{eq:result:pi0:3}) more explicitly, 
the external momenta flowing into
the neutral pion contribution
$\Pi^{\nu_{(1)} \nu_{(2)} \nu_{(3)} \nu}
  \left(k_{(1)},\,k_{(2)},\,k_{(3)}\right)_{\pi^0}$
must obey one kinematic constraint
depending on each of three channels.
 Thus, 
neither Eq.~(19) in Ref.~\cite{Colangelo:2017urn}
nor the well-known formula derived using 
Gegenbauer polynomials in Ref.~\cite{Knecht:2001qf}
can be used for the neutral pion contribution to 
the muon $g-2$, 
because they were derived on the assumption that
{\it 
all components of two loop momenta are completely independent}.
 The formula in Ref.~\cite{Colangelo:2017urn}
%refer to the real part of 
%the amplitude
%$\Pi^{\nu_{(1)} \nu_{(2)} \nu_{(3)} \nu}
%  \left(k_{(1)},\,k_{(2)},\,k_{(3)}\right)$
%(so it 
helps to find the hadronic light-by-light scattering contribution
to the muon $g-2$ 
induced by the states
each of which is composed of two or more hadrons.
\item
\ After the magnetic projection,
the integrand of the contribution to the muon $g-2$ 
consists of the scalar product(s) of loop momenta,  
$l_{(1)}$, $l_{(2)}$, and $p$ ($p^2 = m_\mu^2$)
and the transition form factors.
 The integral on the time component $l_{(1)}^0$ ($l_{(2)}^0$) 
leaves only the terms that have {\it even} power of 
$l_{(1)}^0$ ($l_{(2)}^0$).
 Hence, under Wick-rotation, 
$l_{(1)}^0$ and $l_{(2)}^0$ contained in the scalar products
never multiply the imaginary unit ($\iu \coloneq \sqrt{-1}$) 
to the integrand.
%whether 
%the parts other than the imaginary unit ($\iu \coloneq \sqrt{-1}$) times 
%the product of two pion form factors
%in the expression for the contribition to the muon $g-2$
%are real or pure imaginary.
 {\it It is just the integration measure
$\int d l_{(1)}^0 \int d l_{(2)}^0$ 
that determines
which of the real or imaginary part of the integrand 
is relevant to the muon $g-2$ after Wick-rotation}.
\item
\ In the case that two loop momenta are independent,
both of two time components must be Wick-rotated\,: 
\begin{align}
 &
 \int d l_{(1)}^0 \int d l_{(2)}^0\,g(l_{(1)}^0,\,l_{(2)}^0)
 =
 \int d l_{(1)}^4 \int d l_{(2)}^4\,\iu^2\,
  g(\iu l_{(1)}^4,\,\iu l_{(2)}^4) \,,
\end{align}
yielding the expressions in Ref.~\cite{Colangelo:2017urn,Knecht:2001qf}
for the muon $g-2$. 
 As for the neutral pion contribution to the muon $g-2$ found here, 
the time component of either one of two loop momenta
is not an independent integration variable, 
so that Wick-rotation is performed only for one time component
($q^0$ can also be Wick-rotated) 
\begin{align}
 &
 \int d l_{(1)}^0 \int d l_{(2)}^0\,
  \delta\left(l_{(1)}^0 - l_{(2)}^0 - q^0\right)
  g(l_{(1)}^0,\,l_{(2)}^0)
   \nonumber \\
 &\quad
 =
 \int d l_{(2)}^0\,g(l_{(2)}^0 + q^0,\,l_{(2)}^0)
   \nonumber \\
 &\quad
 =
 \int d l_{(2)}^4\,\iu\,
  g(\iu l_{(2)}^4 + q^0,\,\iu l_{(2)}^4) \,,
\end{align}
where $g(l_{(1)}^0,\,l_{(2)}^0)$ is an even function of $l_{(1)}^0$
as remarked above, 
and every term with odd number of $l_{(2)}^4$ vanishes
under the integral of $l_{(2)}^4$; for instance, 
\begin{align}
 \int d l_{(2)}^4
 \left( \iu l_{(2)}^4 + q^0 \right)^2
 &=
 \int d l_{(2)}^4
 \left\{
  \iu^2 \left( l_{(2)}^4 \right)^2 + \left(q^0\right)^2
 \right\} \,.
\end{align}
\item
\ This way, Wick rotation with respect to only one time-component
will lead the expression picking up only the imaginary part of
$\Pi^{\nu_{(1)} \nu_{(2)} \nu_{(3)} \nu}
  \left(k_{(1)},\,k_{(2)},\,k_{(3)}\right)_{\pi^0}$
to estimate the contribution to the muon $g-2$.
%%%
%\item 
%\ If we can show that the form factor is real, 
%the neutral pion intermediate state contribution 
%$\Pi^{\nu_{(1)} \nu_{(2)} \nu_{(3)} \nu}
%  \left(k_{(1)},\,k_{(2)},\,k_{(3)}\right)_{\pi^0}$
%is pure imaginary.
\end{enumerate}

 Finally, we suggest a new direction to 
improve the estimate of the hadronic light-by-light scattering contribution. 
 As discussed around Eq.~(\ref{eq:ex:insertionOfStates}),
there are other choices for inserting a complete basis 
$\left\{\left|v\right>\right\}$
of QCD states
into the connected Green function of four $J^\mu(x)$.
 The average of those two ways of insertion 
%so as to 
%express each term by the product of 
%the matrix element of a single EM current 
%and the matrix element of the time-ordered product of 
%three EM currents 
leads
\begin{align}
 &
 2\,
 \left< 0 \right|
  \mathscr{T}\left[
   J^{\nu_{(1)}}(z_{(1)})\,J^{\nu_{(2)}}(z_{(2)})\,J^{\nu_{(3)}}(z_{(3)})\,
   J^{\nu}(0)
 \right]
 \left| 0 \right>
  \nonumber \\
 &\quad
 =
 \ 
 \theta\left(
  \min\left( z_{(1)}^0\,,\,z_{(2)}^0\,,\,z_{(3)}^0 \right) - 0
 \right)
  \nonumber \\
 &\qquad
 \times
 \sum_v
  \left< 0 \right|
   \mathscr{T}
   \left[
    J^{\nu_{(1)}}(z_{(1)})\,J^{\nu_{(2)}}(z_{(2)})\,J^{\nu_{(3)}}(z_{(3)})
   \right]
  \left| v \right>
  \left< v \right| J^{\nu}(0) \left| 0 \right>
  \nonumber \\
 &\quad
 +
 \theta\left(
  0 - \max\left( z_{(1)}^0\,,\,z_{(2)}^0\,,\,z_{(3)}^0 \right)
 \right)
  \nonumber \\
 &\qquad
 \times
 \sum_v
  \left< 0 \right| J^{\nu}(0) \left| v \right>
  \left< v \right|
   \mathscr{T}\left[
    J^{\nu_{(1)}}(z_{(1)})\,J^{\nu_{(2)}}(z_{(2)})\,J^{\nu_{(3)}}(z_{(3)}) 
   \right] 
  \left| 0 \right>
   \nonumber \\
 &
 \quad
 +
 \theta\left(
  \min\left( z_{(2)}^0 \,,\, z_{(3)}^0 \,,\, 0 \right) - z_{(1)}^0
 \right)
  \nonumber \\
 &\qquad
 \times
 \sum_v
  \left< 0 \right|
   \mathscr{T}\left[
    J^{\nu_{(2)}}(z_{(2)})\,J^{\nu_{(3)}}(z_{(3)})\,J^{\nu}(0)
   \right]
  \left| v \right>
  \left< v \right| J^{\nu_{(1)}}(z_{(1)}) \left| 0 \right>
  \nonumber \\
 &\quad
 +
 \theta\left(
  z_{(1)}^0 - \max\left( z_{(2)}^0 \,,\, z_{(3)}^0 \,,\, 0 \right) 
 \right)
  \nonumber \\
 &\qquad
 \times
 \sum_v
  \left< 0 \right| J^{\nu_{(1)}}(z_{(1)}) \left| v \right>
  \left< v \right|
   \mathscr{T}\left[
    J^{\nu_{(2)}}(z_{(2)})\,J^{\nu_{(3)}}(z_{(3)})\,J^{\nu}(0)
   \right]
  \left| 0 \right>
   \nonumber \\
 &
 \quad
 +
 \theta\left(
  \min\left( z_{(3)}^0 \,,\, z_{(1)}^0 \,,\, 0 \right) - z_{(2)}^0
 \right)
  \nonumber \\
 &\qquad
 \times
 \sum_v
  \left< 0 \right|
   \mathscr{T}\left[
    J^{\nu_{(3)}}(z_{(3)})\,J^{\nu_{(1)}}(z_{(1)})\,J^{\nu}(0)
   \right]
  \left| v \right>
  \left< v \right| J^{\nu_{(2)}}(z_{(2)}) \left| 0 \right>
  \nonumber \\
 &\quad
 +
 \theta\left(
  z_{(2)}^0 - \max\left( z_{(3)}^0 \,,\, z_{(1)}^0 \,,\, 0 \right)
 \right)
  \nonumber \\
 &\qquad
 \times
 \sum_v
  \left< 0 \right| J^{\nu_{(2)}}(z_{(2)}) \left| v \right>
  \left< v \right|
   \mathscr{T}\left[
    J^{\nu_{(3)}}(z_{(3)})\,J^{\nu_{(1)}}(z_{(1)})\,J^{\nu}(0)
   \right] 
  \left| 0 \right>
  \nonumber \\
 &
 \quad
 +
 \theta\left(
  \min\left( z_{(1)}^0 \,,\, z_{(2)}^0 \,,\, 0 \right) - z_{(3)}^0
 \right)
  \nonumber \\
 &\qquad
 \times
 \sum_v
  \left< 0 \right|
   \mathscr{T}\left[
    J^{\nu_{(1)}}(z_{(1)})\,J^{\nu_{(1)}}(z_{(2)})\,J^{\nu}(0)
   \right] 
  \left| v \right>
  \left< v \right| J^{\nu_{(3)}}(z_{(3)}) \left| 0 \right>
  \nonumber \\
 &\quad
 +
 \theta\left(
  z_{(3)}^0 - \max\left( z_{(1)}^0 \,,\, z_{(2)}^0 \,,\, 0 \right)
 \right)
  \nonumber \\
 &\qquad
 \times
 \sum_v
  \left< 0 \right| J^{\nu_{(3)}}(z_{(3)}) \left| v \right>
  \left< v \right|
   \mathscr{T}\left[
    J^{\nu_{(1)}}(z_{(1)})\,J^{\nu_{(2)}}(z_{(2)})\,J^{\nu}(0)
   \right] 
  \left| 0 \right>
   \,.
  \label{eq:insert:J-JJJ}
\end{align}
 There is no single neutral pion contribution in this representation.
 The intermediate state of two charged pions
$\left|v\right> = \left| \pi^+(\boldsymbol{h}_{(1)})\,\pi^-(\boldsymbol{h}_{(2)}) \right>$
is expected to contribute to Eq.~(\ref{eq:insert:J-JJJ})
dominantly.
 Eq.~(\ref{eq:insert:J-JJJ}) will serve
a completely independent estimate of 
the hadronic light-by-light scattering contribution to the muon $g-2$
if enough knowledge on
$\left< \pi^+(\boldsymbol{h}_{(1)})\,\pi^-(\boldsymbol{h}_{(2)}) \right|
 \mathscr{T}\left[
  J^{\nu_{(1)}}(z_{(1)})\,J^{\nu_{(2)}}(z_{(2)})\,J^{\nu}(0)
 \right]
\left| 0 \right>$
can be obtained from, say, 
the experiment on the production of 
$\pi^+ \pi^- \gamma^* (\rightarrow \mbox{some particles})$
through two photon process.

\section*{Acknowledgment}
The author thanks the collaboration members in Ref.~\cite{Blum:2023vlm}
for discussion, especially ,N.~Christ for questioning the structure of
Eq.~(\ref{eq:known-pion-pole-contribution}), 
and L.~Jin for suggesting use of time reversal
to express the matrix element with the pion as the final state 
by the form factor $F\left(l_{(1)}^2\,,\,l_{(2)}^2\right)$. 
The author is supported in part 
by JSPS Grant-in-Aid for Scientific Research (C) 20K03926.

\let\doi\relax

%%%%%%
%%%%%%
%%%%%%
%%%%%%
%%%%%%
\end{document}